\documentclass[nofootinbib,twocolumn,letter,preprint,aps,10pt]{revtex4}

\usepackage{amsmath}
\usepackage{graphicx}
\usepackage{subfigure}
\usepackage{ulem}
\usepackage{color}

\newlength{\upit}\upit=0.1truein

\newcommand{\ltappr}{{{\lower4pt\hbox{$<$} } \atop \widetilde{ \ \ \ }}}
\newlength{\bxwidth}\bxwidth=1.5 truein

\newcommand{\co}{{\cal O}}

\newcommand{\tr}{{\hbox{Tr}}}

\newcommand{\rar}{\rightarrow}

\newcommand{\be}{\begin{equation}}
\newcommand{\ee}{\end{equation}}
\newcommand{\bal}{\begin{align}}
\newcommand{\eal}{\end{align}}
\newcommand{\bee}{\begin{eqnarray}}
\newcommand{\eee}{\end{eqnarray}}
\newcommand{\bes}{\begin{subequations}}
\newcommand{\ees}{\end{subequations}}

\def\Re{{\rm Re\,}}
\def\Im{{\rm Im\,}}
\def\w{\omega}
\def\p{\partial}

\def\R{{\cal R}}
\def\a{\alpha}
\def\nn{\nonumber}

\def\b{{\beta}}
\def\a{{\alpha}}
\def\g{{\gamma}}
\def\G{{\Gamma}}
\def\d{{\delta}}
\def\D{{\Delta}}

\def\bs{{\boldsymbol\sigma}}
\def\bk{{\mathbf{k}}}
\def\bk{{\mathbf{k}}}

\def\om{{\omega}}

\newlength{\figwidth}
\newlength{\shift}
\begin{document}

\title{Spin and holographic metals
}

\author{Victor Alexandrov and Piers Coleman }
\affiliation{Center for Materials Theory, Dept of Physics and Astronomy, Rutgers University,
136 Frelinghuysen Road,  Piscataway, NJ 08854, U.S.A.
}

\begin{abstract}
In this paper we discuss two-dimensional holographic metals from a
condensed matter physics perspective.  We examine the spin
structure of the Green's function of the holographic metal,
demonstrating that the excitations of the holographic metal are
``chiral'', lacking the inversion symmetry of a conventional Fermi
surface, with only one spin orientation for each point on the Fermi
surface, aligned parallel to the momentum.  While the presence of a
Kramer's degeneracy across the Fermi surface permits the formation of
a singlet superconductor, it also implies that ferromagnetic spin
fluctuations are absent from the holographic metal, leading to a
complete absence of Pauli paramgnetism. In addition, we show how the
Green's function of the holographic metal can be regarded as a
reflection coefficient in anti-de-Sitter space, relating the ingoing
and outgoing waves created by a particle moving on the external
surface.
\end{abstract}

\maketitle


\section{Introduction} \label{sec:1} The past few years have seen a
tremendous growth of interest in the possible application of
``holographic methods'', developed in the context of String theory, to
Condensed matter physics.  Holography refers to the application of the
Maldacena conjecture \cite{Maldacena:1997re}, which posits that the
boundary physics of Anti-de-Sitter space describes the physics of
strongly interacting field theories in one lower dimension.  The hope
is to use holography to shed light on the universal physics of quantum
critical metals\cite{NFL-Lee,NFL-MIT,NFL-Zaanen}.
This paper studies the spin character of the
holographic metal, showing that its excitations are chiral in
character, behaving as strongly spin-orbit coupled excitations with {no inversion symmetry
and} spin aligned parallel to their momentum (see the end of this section).

Quantum criticality refers to the state of matter at a zero
temperature second-order phase transition. Such phase transitions are
driven by quantum zero-point motion.  In contrast to a classical
critical point, in which the statistical physics is determined by
spatial configurations of the order parameter, that of a quantum
critical point involves configurations in space-time with a diverging
correlation length and a diverging correlation time\cite{sachdevbook,hertz,millis}. There is
particular interest in the quantum criticality that develops in
metals, where dramatic departures from conventional metallic behavior,
described by Landau Fermi liquid theory\cite{questions,gegenwart}, are found to develop. Metals
close to quantum criticality are found to develop a marked
pre-disposition to the development of anisotropic superconductivity
and other novel phases of matter\cite{mathur,broun}. The strange metal phase of the
optimally doped cuprate superconductors is thought by many to be a
dramatic example of such phenomena\cite{broun}.

In quantum mechanics, the partition function can be rewritten as a
Feynman path integral over imaginary time.
\be
Z= {\tr}\left[e^{-\beta  H} \right]=\! \int {\cal D}[{\cal O}]
\exp\left[
-\int_{0}^{\frac{\hbar }{k_{B}T}}d\tau L ({\cal O},\tau ) \right]
\ee
where $L$ is the Lagrangian describing the interacting system and $\tau $ the imaginary time, runs from $0$ to $\hbar / (k_{B}T)$. Inside the path integral, the physical fields ${\cal O}$ are periodic or antiperiodic over this interval. The path integral formulation indicates a new role for temperature: whereas temperature is a tuning parameter at a classical critical point, at a quantum critical point it plays the role of a boundary condition: a boundary condition  in time\cite{questions}. When a classical critical system is placed in a box of finite extent, it acquires the finite correlation length set by the size of the box. In a similar fashion, one  expects that when a quantum critical system with infinite correlation time is warmed to a small finite temperature, the characteristic correlation time becomes the ``Planck time''
\begin{equation}
\tau_{T} \sim \frac{\hbar }{k_{B}T}
\end{equation}
set by the periodic boundary conditions. This ``naive scaling'' predicts that dynamic correlation functions will scale as a function of $E/k_{B}T$.  Neutron scattering measurements of the quantum critical spin correlations in the heavy fermion systems $CeCu_{6-x}Au_{x}$ and $UCu_{5-x}Pd_{x}$\cite{schroeder, aronson} do actually show $E/T$ scaling. The marginal Fermi liquid behavior of the cuprate metals that  develops at optimal doping is also  associated with such scaling. The most direct approach to quantum criticality,  pioneered by Hertz\cite{hertz,millis}, in which a Landau Ginzburg action is studied, adding in the damping effects of the metal. Unfortunately, the Hertz approach  predicts that naive scaling only develops in antiferromagnets below  two spatial dimensions. Today, the origin of $E/T$ scaling in the cuprates and heavy fermion systems, and the many other anomalies that develop at quantum criticality  constitutes an unsolved problem. A variety of novel schemes have been proposed to solve this problem, mostly based on the idea that some kind of local quantum criticality emerges \cite{qmsi,varma}, but at the present time there is not yet an established consensus. The hope is that holography may help.

\subsection*{Holographic approach}
To understand the new approaches, we start with a discussion of the Maldacena conjecture, which proposes that  the partition function of a quantum critical (conformally invariant) system can be re-written as a path integral for a higher dimensional gravity (or string theory) problem. In the ``physical'' system of interest the space-time dimension is $d$ while in the gravity problem there is an extra coordinate $r$ and the space time dimension is $D=d+1$.

The Maldacena conjecture can be written as an identity between the generating functional of a $d$ dimensional conformal field theory, and a $d+1$ dimensional gravity problem, $Z_{\text{CFT}}[j ] = Z_{\text{grav}}[\phi]$
\be \label{intro0}
\left\langle e^{-\int \!\!d^{d}x \ j (x) \, {\cal O } (x)}
\right\rangle_{\text{CFT}}= \int D[\phi] \ e^{- \int dr\int d^{d}x   \mathcal L_{\text grav}[\phi]}
\ee
Here $j (x)$ is a source term coupled to the physical field ${\cal O}
(x)$, corresponding for instance to a quasi-particle. The right hand
side describes the ``Gravity dual'', where the gravity fields
$\phi(x,r)$ must satisfy the boundary condition that they are equal to
the source terms $j(x)$ on the boundary $\lim_{r\rar\infty}\phi
(x,r)=j (x)$.  This condition establishes the
relation between the variables of the d-dimensional field theory
and the d+1 dimensional gravity problem in (\ref{intro0}). The lower dimensional theory is conformally
invariant, which implies that the state is critical in space time, i.e
quantum critical. From a condensed matter perspective, the equality of
the two sides implies that the physics of the quantum critical system
of interest can be mapped onto the surface modes of a higher
dimensional gravity problem.

\begin{figure}[h]
  \includegraphics[width=.30\textwidth]{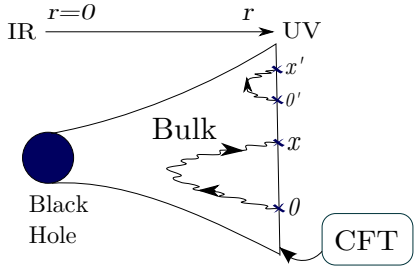}
  \caption{  Illustrating  the surface
  excitations ``propagating'' into the bulk. The  horizontal
  axis is $r$, while the vertical axis is the physical coordinate of the critical theory (CFT). }
  \label{holo}
\end{figure}

A physical picture for the AdS coordinate $r$ is obtained as follows. Consider the injection and removal of a particle on the boundary of the AdS space, separated by a distance $x$, as illustrated in fig.~\ref{holo}. When the point of injection and removal are nearby, the Feynman paths connecting them will cluster near  the boundary, probing {\sl large} values of r. By contrast, when the two points are far apart, the Feynman paths connecting them will pass deep within the gravity well of the Anti-de Sitter space, probing {\sl small} values of $r$ close to the black hole. Hence $r$ represents an energy scale of the problem (corresponding to  the  ultra-violet cut-off
in a renormalization group flow).

\begin{figure}[h]
      \includegraphics[width=.5\textwidth]{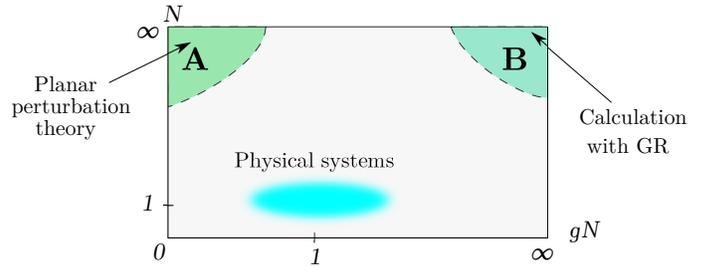}\\
 \caption{Schematic  diagrams for the critical (CFT) theory. Region {\bf A}  can be computed perturbatively on CFT side but is highly non trivial on the string theory side.  This paper is about region {\bf B} where critical theory is strongly correlated but computable with GAR. The real physical models  have only $N=1,2$ and $g\sim1$ and thus are in the center of the diagram.  }
 \label{pic_chart}
\end{figure}

The notion that condensed matter near a quantum critical point
might acquire a simpler description when rewritten as a gravity dual seems at first surprising, especially considering that the  higher dimensional dual is a ``string theory'' of quantum gravity. The essential  simplification occurs in the large $N$ limit.  Here, most of the understanding derives from the particular case where the  Maldacena conjecture has been most extensively studied and  corroborated --  a family of $SU (N)$ supersymmetric QCD models with two expansion parameters: a gauge coupling  constant $g$ and number of gauge fields $N^2$, as summarized in  Fig.~\ref{pic_chart}. The corresponding gravity dual, is a string theory with ``string coupling constant'' $g_{str}$  and a characteristic ratio $l_{str}/L$ between the string length $l_{str}$ and the characteristic length of the  space-time geometry where the string resides.  While $g_{str}$ controls the amplitude for strings to sub-divide, changing the genus of the world-sheet, $l_{str}$ constrains the amplitude of string fluctuations. The correspondence implies
\be g\sim
g_{str}, \quad g N \sim (L/l_{str})^4
\ee

Each point of fig. \ref{pic_chart} has a dual string description.
For large $N$ and small $g$ (region {\bf A}) the critical theory can
be computed in perturbation series but a string description is
extremely complicated. Some have even suggested this might be way of
solving string theory by mapping it onto many body physics \cite{Strominger}.  The focus
of current interest in holographic methods
is on region {\bf B}, in the double limit $g, N\rar
\infty$, that corresponds to $l_{str}\rar 0$ or just classical
gravity. In this sense then, the Maldacena conjecture, if true,
provides a new way to carry out large $N$ expansions for quantum
critical systems. Since we don't yet have a working large $N$ theory
for quantum critical metals, this may be a useful way of
proceeding. A similar philosophy has also been applied
in the context of nuclear physics, as a way to place a
theoretical limit on the viscosity of quark gluon plasmas
\cite{Son:2007vk}.

The field is at an extraordinary juncture. On the one hand, it is
still not known whether the Maldacena conjecture works for a much
broader class of models, yet on the other, the assumption that it does
so, has led to an impressive initial set of results. In particular, a
charged black hole in Anti-de Sitter space appears to generate a
strange metal \cite{NFL-MIT,NFL-Lee,NFL-Zaanen}, with a Fermi surface
at the boundary of the space and novel anomalous exponents in the
self-energy.  A fascinating array of results for the strange metals
have been obtained, including the demonstration of singlet
pairing\cite{Faulkner:2009am}
 and even the development of de Haas van Alphen
oscillations in the magnetization in an
applied field\cite{Denef:2009yy}.

\subsection*{Motivation and results}

This paper describes our efforts to understand the ramifications of these developments. One of the motivating ideas was to develop a better physical picture of the strange metal. We were particularly fascinated by the attempt to describe high T$_c$ superconductivity \cite{Hartnoll:2008vx, HartnollBCS} (see \cite{Horowitz:2010gk} for review): in the presence of a charge condensate in the bulk, the boundary strange metal develops a singlet s-wave pair condensate \cite{Faulkner:2009am}.
The formation of singlet s-wave pairs indicates that the strange fermions carry spin, motivating us to ask whether there is a paramagnetic spin susceptibility associated with the strange metal. This led us to examine the matrix spin-structure of fermion propagating in the strange metal.

Spin is a fundamentally three dimensional property of non-relativistic electrons, and in the absence of spin-orbit coupling it  completely decouples from the kinetic degrees of freedom as an independent degree of freedom, a common situation in condensed matter physics. By contrast, in the holographic metals studied to date, the particles are intrinsically two dimensional. For these particles, derived from {\sl two component} relativistic electron spinors, there is no spin. One way to see this is to look at two components of the fermion, which describe the electron and positron fields in two dimensions, leaving no room for spin. How then is it possible to form a spin-singlet superconductor from these fields, when there is no spin to form the singlet?

In this paper, by examining the spin structure of holographic metals we  contrast some  important similarities and differences between holographic metals and real electron fluids.
In our work we have two main results:
\begin{enumerate}

\item We show that the excitations of the strange metal are
chiral\footnote{\label{footnote2} Here we use ``chirality'' in the
sense adopted by condensed matter physics, to mean the helicity or
handedness of a particle.} fermions, with spins orientated parallel to
the particle momenta. Near the FS the Green's function becomes
    \begin{equation}\label{intro2}
    G_{ w\rar 0 } \ \ =  {Z(w)\over  \om - v_F\bs\cdot \bk + \Sigma (\om)} + G_\text{incoh}
    \end{equation}
The strong spin-momentum coupling generated by the term $\bs\cdot
\bk$ means that the Fermi surface preserves time-reversal
symmetry, but violates inversion symmetry. In particular,
a simple spin reversal at the Fermi surface costs an energy
$2v_{F}k_{F}$, so that the spins are preferentially aligned parallel
to the momenta to form chiral fermions. In this way, spin ceases to
exist as an independent degree of freedom in two-dimensional
holographic metals, as opposed to a spin degenerate interpretation
(\ref{GF:polchinski}). One of the immediate consequences of this
result is that the most elementary property of metals, a Pauli
susceptibility, is absent.

\item We identify an alternate  interpretation of the holographic Green's functions\footnote{\label{footnote1}We only use $G$ do denote the retarded Green's function.} as the reflection coefficient of waves emitted into the interior of the Anti de Sitter space by the boundary particles, as they reflect off the  black hole inside the anti-de Sitter bulk. Namely
    \begin{equation}\label{intro1}
     G = M_k \R(\om,\bk)
    \end{equation}
    where $\R$ is the reflection coefficient associated with the  black hole and $M_k$ is a known kinetic coefficient. For bosons $M_k=1$ while for  fermions $M_k=M(\w,\bk)$ has more involved structure (\ref{ferm:GF_refl}). The reflection $\R$ contains  the information about the branch cuts and excitation spectra.

\end{enumerate}
We discuss the full implications of these results in the last section.

\section{Background Formalism}

Our goal is to  determine the holographic Green's functions using
linear response theory. Here, for completeness we provide  some of the
background formal development\footnote{Throughout the paper all the
quantities are dimensionless including  e.g. temperature.}.
For details, we refer the reader to extensive reviews\cite{Review:Hartnoll,Review:McGreevy,Review:Liu,Review:Sachdev}.

 The main conjecture \cite{Maldacena:1997re} connecting currents $j$ in lower dimensional CFT and fields of the bulk gravity (as a limit from string theory)
\be
\label{conjecture}
Z_{\text{CFT}}[j ] = Z_{\text{grav}} [\phi]
\ee
 where $\phi$ and $j$ are related by the boundary condition
\be
 j(x)=\lim_{r\rightarrow \infty }\phi (r;x) r^{d-\D}\label{conjecture2}
\ee
the power of $r$ reflects the scaling dimension of the source dim$[j]=\D-d$. The source is coupled to the physical field, better thought as quasi particle, denoted by ${\cal O} (x)$. $\D$ is the conformal dimension of that field dim$[\co]=\D$, namely
  \be \nn
Z_{\text{CFT}}[j ]=
\left\langle \exp \left[\int d^{d}x j (x){\cal O }(x)
\right]\right\rangle_{CFT}.
\ee
 This generating functional  determines the physics of the quantum system. The gravity part can be computed classically
  \be
  Z_{\text{grav}} = e^{-S_{\text{grav}}},
  \ee
derivatives of the generating functional $Z[j]$ determine the Green's functions of the fields ${\cal O} $
\be
    \label{scal:vev}
    \langle \co \rangle \equiv   \left.{\d Z[j] \over \d j } \right|_{j=0} =\lim_{r\rightarrow \infty} {  r^{\D-d } } {\d S_{\rm grav}[\phi] \over \d \phi},
\ee
The holographic Green's functions can be obtain from the  quadratic components of the action. The equation of motion then has two independent solutions near the boundary
\begin{align} \label{scal:asympt.general}
\phi =   \underset{\text{\normalsize in-going}}{A r^{\D-d}} +       \underset{\text{\normalsize out-going}}{ B r^{-\D}}       + \  ... \ , \qquad \text{as } r\rar\infty
\end{align}
Usually  the ingoing component  $A$ is  referred as the ``non-normalizable" mode, while the outgoing component $B$ is the ``normalizable" mode. Note how the exponents of $r$ match the dimensions of the source and the {response} $\co$. In the absence of the source term $j$, the solution must vanish at infinity and the outgoing component vanishes.  Once we turn on the source $j$, the Maldacena condition  (\ref{conjecture2}) that $\phi  (r,x)\rightarrow j (x)$ enables us to identify $A$ as the source
\be\nn
j\equiv   A.
\ee
Accordingly, the  outgoing mode corresponds to the response\footnote{indeed, after substituting (\ref{scal:asympt.general}) into (\ref{scal:vev}) and varying it w.r.t. source $A$   (consequently setting source to zero) we are left with the term proportional to $B$.}~$\langle \co\rangle$
\be\nn
\langle \co \rangle =  const  \cdot B
\ee
up to a numerical constant dependent on the particular theory at hand. For a free scalar  $const= {(2\D - d})$, for a fermion  $const= i \g^t$. In {a} systematic treatment one needs to regulate the procedure by adding boundary terms, (see  appendix \ref{boundary.terms}).

Since the Green's function is the linear response to the source, it follows that up to a constant of proportionality
\be
G=const \cdot  B/A.
 \ee

The procedure to extract the Green's function of a holographic metal is then:
  \begin{enumerate}

  \item Select a background allowing black hole and usually asymptotically AdS.

  \item Select {the } bulk field content and Lagrangian.

  \item Select one of the fields with the quantum numbers (spin, charge, etc) of the desired operator $\co $.

  \item Solve  the classical field equations in that background, including {the} backreaction on the gravity.

  \item Find the  asymptotics of the fields at the boundary. Find the $\D,$ outgoing (leading) and ingoing terms by comparing with (\ref{scal:asympt.general}).

  \item The ingoing amplitude at the boundary represents the source, the outgoing amplitude gives the response, the Green's function is the ratio of the two.

 \end{enumerate}

\subsubsection{Examples}
We now sketch these steps for the
scalar and fermion cases.
The first step is to choose a background. One of the well known solutions of Einstein-Maxwell equations is the Reissner-Nordstr\"om (RN) black hole. This background involves a nontrivial electric field ($E_r=-\p_r A_0 $) and asymptotically AdS metric $g_{\mu\nu}$. In the units where horizon $r=1$, the metric, fields and temperature $T$ are
\begin{eqnarray}\label{RN}
&ds^2  =  {r^2} (-f dt^2 + d x_i^2)  +  {1\over r^2 f}dr^2, \\\label{grav:def2}
&f=1-{Q^2 +1\over r^3}+{Q^2\over r^4}, \\\label{RN3}& T = {3-Q^2\over 4\pi},
   \qquad  A_0 = \mu \left(1- {1 \over  r}\right).
\end{eqnarray}
Alternate solutions to the metric differ only in the profile function ("blackening factor") $f(r)$, and the horizons are defined by the zeros of $f(r)$ (as one approaches the horizon time coordinate becomes irrelevant). The solution (\ref{RN}) describes a black hole with electric charge $Q$ in a space with negative cosmological constant. The negative  cosmological constant causes the space-time to be  asymptotically AdS and thus to have a boundary. The scalar potential $A_0$  at the boundary goes to a constant $\mu$, the  chemical potential of the boundary theory. Indeed, the  RN black hole is a result of steps 1-4 for just one extra field in the bulk, gauge field $A^\mu$, which is conjugate to the charge currant operator $\mathcal{J}^\mu$. A non-vanishing $A_0$ then  corresponds to a finite source for $\mathcal{J}^0$, which is in fact, the chemical potential $\mu \mathcal{J}^0$.

$\bullet$   \textbf{Bosons}. We choose a bulk action
\be\label{scal:action}
S=S_{GR} + S_{EM} +\int d^4x (-|\p \varphi|^2 - m^2 |\varphi|^2 )
\ee
and the boundary term for a stable solution
\be\label{scal:action.bnd}
S_{bnd}=(\D- d) \int_\p d^3x  \ |\varphi|^2,
\ee
where $S_{GR} +S_{EM}$ is Einstein-Maxwell action. (The term $m^2$ can sometimes be slightly negative\footnote{So called BF bound \cite{BF} $m^2 \geq - 9/4$ for AdS$_4$ with radius $L=1$.}). In the relativistically invariant measure $d^dx$ we  have omitted the factor $\sqrt{g}$, where $g$ is the  determinant of the metric. This action implies the Einstein-Maxwell equations (solved by the RN black hole background) and free scalar equation. For the boundary terms see {the} appendix. The Klein Gordon equation in curved space is then
\begin{equation}\label{scal:eom}
    (D_\mu^2 +m^2)\varphi=0,
\end{equation}
where
\bee
\nn D^2_\mu\equiv D_\mu D^\mu =  (\nabla_\mu - iqA_\mu)(\nabla^\mu - iqA^\mu) =\\= \nabla_\mu\nabla^\mu-iq\nabla_\mu A^\mu - 2 i q A^\mu \nabla_\mu - q^2 A^\mu A_\mu.
\eee
Here, the covariant derivative  $\nabla_\mu$ is defined in terms of the metric, for instance  $\nabla^\mu\nabla_\mu \phi= {1\over
\sqrt{g}}\p_\mu(\sqrt{g}\,\p^{\mu}\phi)$, metric $g$ is given in
Equation (\ref{RN}). Using a little
general relativity and the  Fourier transformed $\varphi = \phi\  e^{-iwt +ikx}$ one can write (\ref{scal:eom}) as (m=0)
\be \label{EOM scalar 2}
\phi'' + {(r^4f)'\over r^4 f}\, \phi' + { (\w+qA_0)^2 -fk^2 \over  r^4 f^2}\, \phi = 0.
\ee
Now we are to solve the  equation to find the asymptotics and  identify the ingoing and outgoing modes. Since it is a second order differential  equation, the full solution can be found numerically, but {the} asymptotics at $r \rar \infty$ are easily
extracted analytically, using $f(r)\rar 1$.
\be \label{EOM scalar 2}
\phi'' + {4\over r}\, \phi' = 0
\ee
hence
$$\phi = A + Br^{-3}.$$

{The} leading ingoing  term is {a} constant $A$, hence $\D =
3$,
{while} the outgoing term should be $r^{-\D}$, cf. (\ref{scal:asympt.general}). To get the proportionality constant we use (\ref{scal:vev}).
  \be
 \langle \co \rangle = - 3  B, \ee
 so the
retarded Green's function is
 \be\quad G=- 3 B/A \label{scal:gf}
\ee
which is actually a $m=0$ case for $G  = (2\D - d) B/A$.

$\bullet$   \textbf{Fermions}. We can introduce the action
\be \label{ferm:action}
S=S_{GR} + S_{EM} +S_{\varphi} + \int d^{4} x \  i\  \overline{\psi} (\g^\mu D_\mu - m) \psi
\ee
with $\overline{\psi} =\psi^\dag \g^t$, and the boundary action
\be  \label{ferm:action.bnd}
S_{bnd}=\int_{\p}  d^3x  \ \overline{\psi} \psi.
\ee
The main difference now is the fact that {a} fermion in 4 dimensions has four components: four quantum-mechanical degrees of freedom (simply spin up, spin down, electron, positron) but the boundary fermion has only 2 components. Thus the other  half of the components should not play a role.  Without loss of generality {the} mass m is positive. In the addition to {the} Einstein-Maxwell {equation} this action also implies {the} Dirac equation
\be\label{ferm:eom}(\g^\mu D_\mu -m)\psi = 0\ee

Before the next step we rewrite (\ref{ferm:eom}) in its full form:
\begin{eqnarray}\label{ferm:dirac}
     (\sqrt{g^{rr}}\G^r\p_r -i\sqrt{g^{00}}\G^0 w + i \sqrt{g^{ii} }\G^i k_i -m)\psi =0 \\
     \text{with:} \quad w = \w +q A_0 \nn
\end{eqnarray}
As in the scalar case, to determine Green's function, we examine the asymptotics boundary behavior, $r\rar \infty$.
In our basis\footnote
                {
                \label{gamma_matr} In a special choice of basis,
                $\Gamma^r =\left( \begin{array}{cc}
                1 & 0  \\
                0 & -1
                \end{array} \right)$,
                $\Gamma^\mu = \left( \begin{array}{cc}
                0 & \gamma^\mu  \\
                \gamma^\mu & 0
                \end{array}
                \right) $,
                where $\g^0 = i\sigma^3$, $\g^x = \sigma^2$,
                $\g^2=-\sigma^1$. and
                \be
                \label{ferm:scaling}
                \psi =\left(\begin{array}{c}
                         \psi_+ \\
                         \psi_-
                       \end{array}\right) \left(-{g_{rr} \over g}\right)^{-1/4} e^{-i\omega t +i k_ix^i}
                       \ee
                       here $g_{\a\b}$ is a metric of a space-time and  $g$ is its determinant
                       }
\be \label{ferm:dirac_x}
\begin{pmatrix} m+ r\p_r & i { \g^\mu k_\mu\over r} \\ i{ \g^\mu k_\mu\over r} &m- r\p_r \end{pmatrix}{\psi_+\choose \psi_-}=0,
\ee
 here $k^\mu=(w,\bk)$.
This is {a} $4\times4$ matrix equation with the following solution.
\begin{equation}\label{ferm:asympt}
 \psi \rar r^{-\frac 32}\left(\begin{array}{c}
         \psi_+ \\
         \psi_-
       \end{array}\right) ;\quad    \left\{\begin{array}{c}
         \psi_+  = A r^m +D r^{-m-1}\\
         \psi_- =B r^{-m} + C r^{m-1}
       \end{array}\right.
\end{equation}
From the four terms we choose in- and out-going modes {in an} analogous {fashion} to the scalar case, the leading term $A r^{m-3/2}$ is in-going, while the term  $A r^{-m-3/2}$ is outgoing and the dimension is
\be
\D =m+3/2.
\ee
The other terms ($C,D$)  are related to the first two. To {determine this dependence} we need to substitute the asymptotics back into the Dirac equation (\ref{ferm:eom}), which leads to
\be
 B= {2m +1 \over i \g^\mu k_\mu} D.
\label{ferm:mk}\ee
 As in (\ref{scal:gf}) the Green's function can be expressed in terms of $A$'s and $B$'s (note $A$ and $B$ are spinors):
\be
       \label{GF:diagonal.prescription}
  G  A =i \g^tB.
\ee
As before, we  obtain the coefficient $i \g^t$ by taking a derivative of the action and $\g^t$ is a part of {the} definition of $ \overline{\psi} $. Armed with (\ref{ferm:mk})
\be  \label{GF:mk}
  G  A = \g^t  {2m +1 \over \g^\mu k_\mu} D \equiv M_k D.
\ee

\section{Reflection Approach} \label{sec:3}

In the previous section we  made use of the ingoing/outgoing
terminology. In this section we identify these modes explicitly.
Here we show how to redefine the problem in a form  resembling the quantum mechanics of a reflected wave. This can be done by transforming coordinates according to $d s = F(r)dr$ and rescaling the wave function as $\psi(r) = Z(r) \Psi(s) $. The original equation then becomes a zero energy scattering problem
\be
 (\p_s^2 - V)\Psi(s)=0,
  \ee
 with a complicated and non-unique function $V(s)$. (The retarded Green's function must be derived using an infalling boundary condition at the black hole horizon \cite{Son:2002sd}. An  advantage of this approach is that the infalling wave  condition  is derived as immediate consequence of the scattering problem.)

  \textbf{Bosons.} Consider again the scalar probe Equation of motion \eqref{EOM scalar 2}  in the RN black hole geometry, defined in (\ref{RN}-\ref{RN3}). The generalization to the full backreacting solution is straightforward.   The rescaling of  coordinates and fields $\varphi(r) \rightarrow \phi(s)$
\begin{equation}\label{refl:transformation}
     ds=Fdr ,\ \ \ \ \varphi = Z \phi
\end{equation}
leads to the following
\begin{gather} \nn
  (\p_r^2 +D_1\p_r +D_0)\varphi \rar \\
   (\p_s^2 +{ D_1 +\frac{F'}{F} +\frac{2Z'}{Z}\over F}\p_s + {D_0 + \frac{Z''}{Z} +D_1 \frac{Z'}{Z}\over {F^2}})\phi =0. \nn
\end{gather}
 There are {an} infinite number  of ways to tune equation (\ref{EOM scalar 2}) to the form of the Schrodinger equation by canceling $\p_s\phi$ term. We will choose the {\it non-unique} combination
 \be
  F= i /fr, \qquad Z=1/r^{3/2}.
 \ee
 This leads to the zero energy scattering problem  $\p_s^2 \phi -V(s)\phi = 0$ with potential given by
 \begin{equation}\label{refl:potential}
    V(r) = {w^2 -k^2f-m^2fr^2\over r^2} - f^2 \left(\frac94 + \frac{3f'}{2rf} \right)
 \end{equation}
 It is useful to write  the limiting values of this potential. It turns out that it goes to a constant on both  the horizon and the boundary.
 \be
 V(r\rightarrow \infty) = -m^2- {9/4},\quad V(r\rightarrow \text{horizon}) = \omega^2
  \ee
     \begin{figure}\begin{center}
      \includegraphics[width=.45\textwidth]{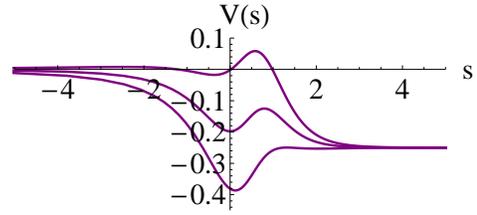}\\
    \caption{Typical Schrodinger potential  $ m^2=-2, \ \ Q^2 = 3, \ k={1,1.6,2} ,\ \omega=0$ (from the top). The incoming wave propagates from the right with zero energy.}\label{schroedinger}
\end{center}\end{figure}
 Finally, we reformulate the Green's Function as a reflection coefficient. Namely
 \be\label{scal:refl}\phi(r\rightarrow \infty) = A \left(  e^{i s (\Delta-3/2)} + \R  e^{ -i s (\Delta-3/2)} \right)
 \ee
  leading to
 \be
 G(\omega, k) = \R
 \ee

  \textbf{Fermions}  in this case the reflection coefficient becomes a matrix. We have already {written} the Dirac equation in the black hole background  (\ref{ferm:dirac_x}). One  can "square' the first order $4\times4$ matrix equation to obtain {a} second order $2\times2$ equation. After rescaling the fields in the same fashion as in the scalar case, the potential acquires the same form as in the scalar case in fig. \ref{schroedinger}, {but} with the different limits
 \be
 V(r\rar  \infty) = -(m+1/2)^2,\quad V(r\rightarrow \text{horiz.}) = \omega^2.
  \ee
As in (\ref{scal:refl}) the solution is a superposition of incident and reflected waves:
 \begin{equation}\label{ferm:asympt2}
   \psi_+  \rar  \left(e^{i s (m+\frac12)} +\R  e^{-i s (m+\frac12)}\right) A,
 \end{equation}
where the reflection coefficient $\R$ is now a \emph{matrix}. The Green's
function is proportional to $\R$ up to a kinematic factor $M_k$
\begin{equation}\label{ferm:GF_refl}
    G = M_k \R =  {2m +1 \over (\omega+q\mu) - \bs \cdot \bk } \R.
\end{equation}
Here we use equation (\ref{GF:mk}). However suggestive the form of $M_k$ is, its poles do not affect  {the} physical non-analyticities of the Green's function G: it is  $\R$ which contains all the relevant poles and branch cuts.


\section{Spin structure} \label{sec:4}

We now return to the question of {the} spin character of the
holographic fermion.  In quantum critical metals, the spin degree of
freedom plays an essential role.  For example, the application of a
magnetic field, via the Zeeman coupling, allows one to tune the system
through a quantum critical point. The presence of critical spin
fluctuations is thought to play an important role in break-down of
Landau Fermi liquid behavior.  This then raises the question as to
whether the holographic fermions obtained by projection from
four-dimensional anti-de-Sitter space, carry a spin quantum
number. Furthermore, what is the nature of the soft modes that drive
the quantum criticality, and is it possible to gap these modes,
driving a transition back into a Fermi liquid?  The boundary fermions
that form about a $D=4$ anti-de-Sitter space are Dirac fermions
described by a two component spinor. Hartnoll at al
\cite{Hartnoll:2008vx} have shown that when a condensed Bose field is
introduced into the bulk gravity dual, these Fermi fields form s-wave,
singlet pairs.  This establishes that the fermions do indeed carry
spin, however, as we shall now show, this spin is
``chiral''${}^\text{\ref{footnote2}}$ and is aligned rigidly with the
momentum of the excitation at the Fermi surface.  There is no
inversion symmetry, and at each point on the Fermi surface there is a
single spin polarization. However, time reversal is not broken, and
reversing the spin also implies reversing the momentum, so it is still
possible to form pairs by combining fermions with opposite spin on
opposite sides of the two dimensional Fermi surface.

It is often tacitly  assumed  that the excitations of holographic metals are {\it non-}relativistic fermions, with an independent spin degree of freedom. In this case the Green's function $G_{\a\b} = \d_{\a\b} G$  would be  proportional to the unit operator as in \cite{Polchinski}\footnote{The effective model of "Semi-Holographic Fermi liquid" was proposed, with lagrangian $\mathcal L= i [
c_{k,\a}^\dag (\omega-\epsilon_k +\mu)c_{k,\a}  +\chi^\dag \Sigma^{-1} \chi_\a+g c_{k,a}^\dag \chi_\a +h.c.
]$ leading to the Green's function (\ref{GF:polchinski}) degenerate in $\a$.}

\be\label{GF:polchinski}
G_{\a\b} =\d_{\a\b}{1\over  \omega- v_F (k-k_F) -g^2 \Sigma}
\ee

Here we shall argue that this is not the case and
 the spin-orbit coupling  remains very large in holographic metals despite the formation of a Fermi surface,  forcing the spin to align with the momentum.

\begin{figure}[h]

\setlength{\unitlength}{1mm}
\begin{picture}(80,45)
\put(30,-10){\includegraphics[width=.3\textwidth]{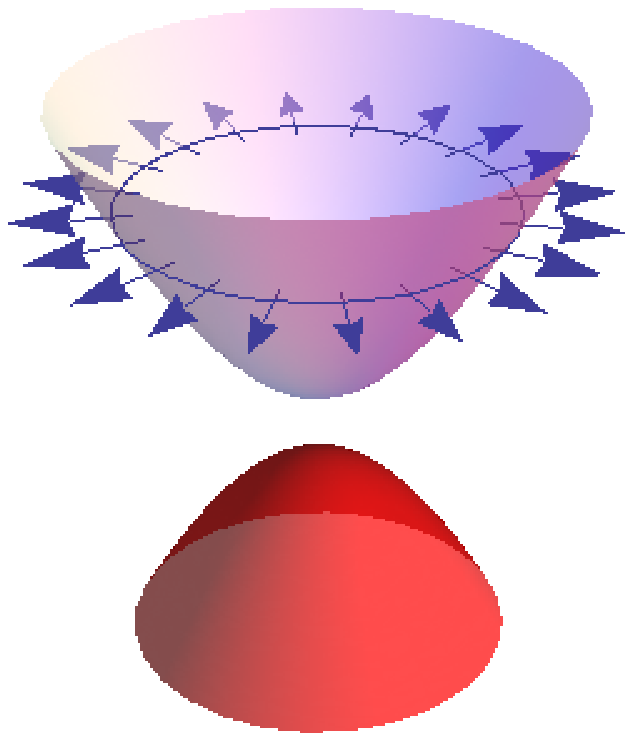}}
\put(-5,-7){\includegraphics[width=.26\textwidth]{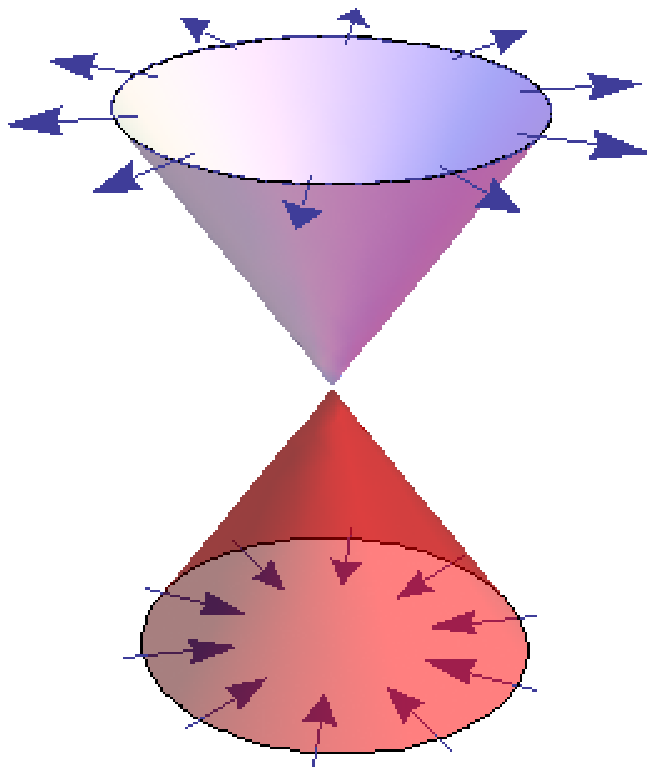}}
  \put(18.4,8){\vector(0,1){35}}
 \put(20,41){Energy}
\put(5,40){$(a)$}
\put(40,40){$(b)$}

 \end{picture}

    \caption{Typical dispersion and Fermi surface of Dirac  fermion in 2+1 (a); and  the holographic metal from fig.\ref{pic_g11re}-a (b). }\label{pic_fs}
\end{figure}

First consider the relativistic case without the black hole when the surface excitations are undoped and form a strongly interacting Dirac cone of excitations with Lorentz invariance. The corresponding Lorentz invariant correlation function is $ \langle  \co  \bar{\co} \rangle ^{-1} =  \tilde{C} k_\mu\g^\mu$, where $ \bar{\co} =\co^\dag \g^0$ and $\tilde{C}$ is an arbitrary function  of 3-momentum $k =\sqrt{\omega^{2}-\bk^{2}}$ . For non-relativistic applications we are  interested in $\langle  \co  \co^\dag \rangle$ and we turn to  a Hamiltonian formalism, treating time and space separately.
The Green's functions takes the form $G^{Lor}=\langle  \co  \co^\dag \rangle= -\langle  \co  \bar{\co} \rangle \g^0 $
\be \label{GF:lor}
G^{Lor}(\w,\bk) =  [\tilde{C} (k) (\bk\cdot\bs -\om) ]^{-1}.
\ee
For the case of  zero bulk fermion mass $\tilde{C} (k) = 1/k$.
 Here we have introduced the Pauli matrixes $\sigma^{i} = \g^{i}\g^0$, for $i=1,2$. Eq. (\ref{GF:lor}) describes two Dirac cones as depicted on fig.\ref{pic_fs}-a, where upper and lower cones have the opposite chirality. There is only one spin orientation parallel to the momentum at any given energy.

Once we add a charged black-hole, the boundary excitations are
``doped'', acquiring a finite Fermi surface (and Fermi velocity $v_F$)
that breaks the Lorentz invariance down to a simple rotational
invariance.
The only rotationally invariant way in which spin can enter, is in the form of a scalar product
with the momentum $\bk \cdot \bs $.  The most general Green's function now takes the
form
\begin{equation} \label{GF:ansatz}
    G (\omega,\bk )  = [C_1 (\omega,k) \bk \cdot \bs + C_2 (\omega,k)]^{-1}
\end{equation}
where $C_{1}$ and $C_{2} $ are two arbitrary functions which
depend on the frequency $\omega$ and the magnitude of the
non-relativistic momentum $k=\vert \bk \vert $.

The physical properties of the theory depend on the form of the coefficients $C_{1,2}$. For example, if $C_1=0$, then
the Fermi surface would be spin degenerate ($\bs$ independent). If $C_1$ is finite and purely real, the momentum becomes strongly coupled to the spin via $ \bk \cdot \bs$ (spin flip does change the ground state) and we are dealing with chiral excitations. One can interpret  $C_1$
as a wave-function renormalization: $C_1=v_F Z^{-1}$ and $C_2$ as a
self energy: $C_2= Z^{-1}(\om +\Sigma - \mu )$.  However, if $C_1$ has
an imaginary part, while the chiral property remains, spin flips  become highly incoherent in nature,  so a standard decomposition of the quasiparticle along the lines of the electron phonon problem is not possible.

To  bring out the chiral ($C_1\neq0$) properties we introduce the chirality projection operators
\begin{equation}\label{}
\Pi_\pm = \frac12 \left(1\pm {\bk \cdot \bs\over k}\right),
\end{equation}
which leads to
  \be \label{GF:ansatz3}
  G=\Pi_+G_{11} +\Pi_- G_{22},
    \ee
where $G_{11}$ and $G_{22}$ are the eigenvalues of the matrix $G$ and    $ 2k C_1 = G_{11}^{-1}-G_{22}^{-1}$, $
    2C_2 =  G_{11}^{-1}+G_{22}^{-1}$.

  \begin{figure}[h]\begin{center}
\includegraphics[width=.24\textwidth]{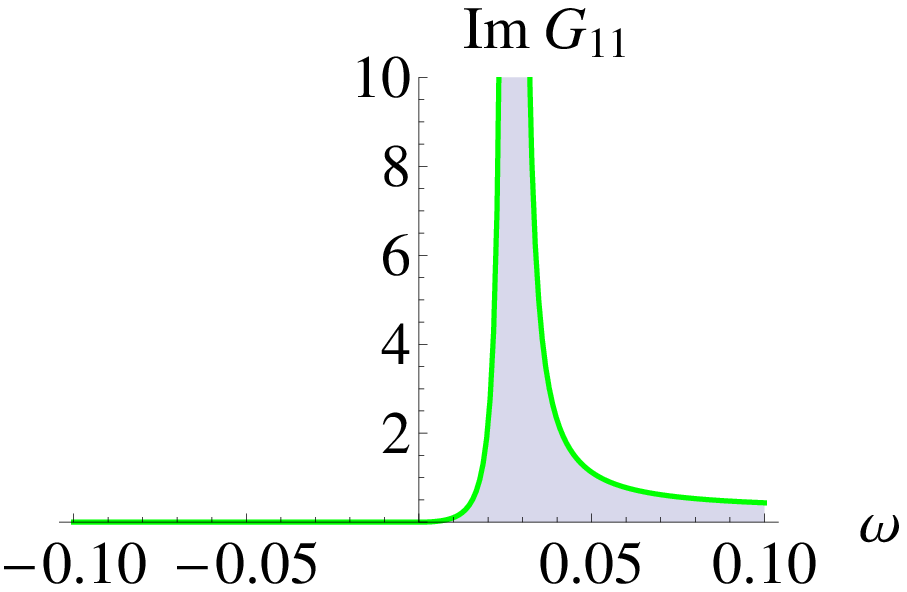}~
\includegraphics[width=.24\textwidth]{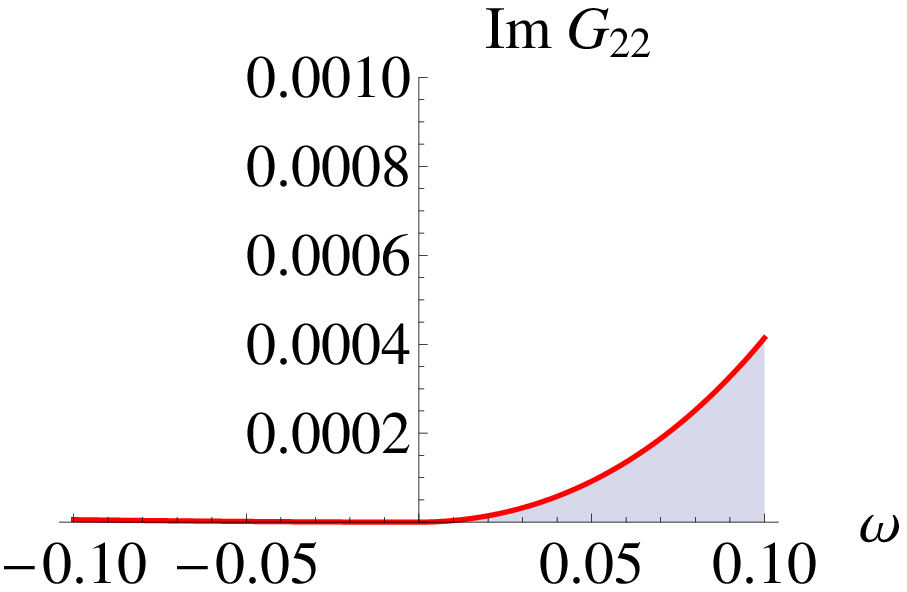}
    \caption{Typical behavior of a spectral function  $\pi A(\om,k) = \Im G_{11} +\Im  G_{22}$ at fixed momentum $k =k_F + 10^{-2}$. $\Im G_{11}$ has a spike that sharpens for $k\rar k_F$ and $\Im G_{22}$ is suppressed. Snapshot for RN black hole  with $q=2$, $m=-1/5$. }\label{pic_g11}
\end{center}\end{figure}
Fig. \ref{pic_g11} illustrates  a typical numerical solution for the eigenvalues.  $\Im {G_{ii}}$ is plotted for fixed $k$ close to the Fermi surface. One eigenvalue has a peak, which we interpret as the chiral quasiparticle component to the spectral function while the other, corresponding to the incoherent background created by flipping the spin anti-parallel to the momentum, lacks any sharp features and \textit{goes to zero} at the Fermi surface. The spectral function $A(\om,\bk) = \frac{1}{\pi}\Im  \tr G$ is${}^\text{\ref{footnote1}}$

\begin{eqnarray} \label{GF:spectral}
  A(\om,\bk) = \frac{1}{\pi}\Im  [G_{\text{chiral}}(\w,\bk)  + G_{\text{incoh}}(\w,\bk) ],
  \end{eqnarray}
where $G_{\text{chiral}}$ refers to the coherent part of the resonance and $ G_{\text{incoh}}$ is the incoherent background.

To further emphasize the chiral structure we use an analytic form of the Green's function. Near $k=k_F$ and $\omega\rightarrow 0$ this can be obtained \cite{Faulkner:2009wj} by matching {the} infra-red 'inner'  region of the RN black hole to the asymptotic AdS 'outer' region.
\be
G_{11}= {z \over\om  - v_F( k-k_F)  + c_1 \om^{2\nu}}  ,
\ee
where $z,v_F$ are real and $c_1$ is a complex constant.
The exponent $\nu \equiv \sqrt{m^2+k_F^2 -q^2/2}/\sqrt{6}$ depends on the mass and the charge of the fermion. Armed with (\ref{GF:ansatz3}) we arrive to
 \be
G\ = {z \over  \w - v_F (\bk \cdot \bs -k_F)  + c_1 \om^{2\nu}}  + G_{incoh}
\ee

\begin{figure}[h]\begin{center}\hspace{-.5cm}
\subfigure[\ $m=0$]{
\includegraphics[height=45mm]{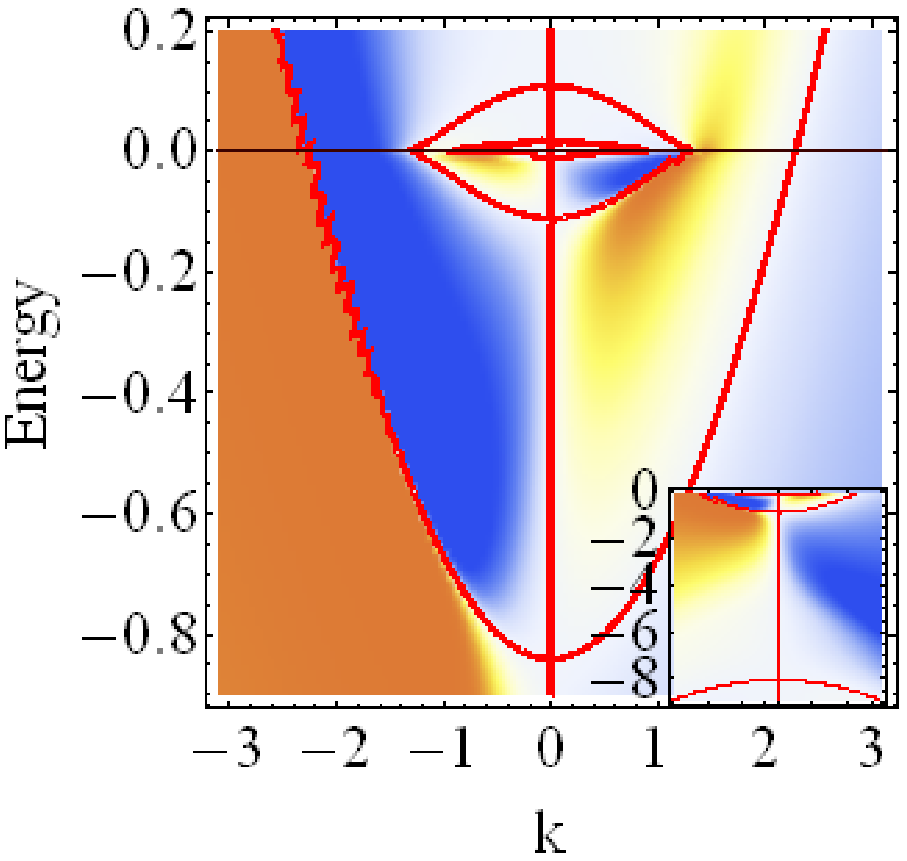}
}~\hspace{-.3cm}
\subfigure[ \ $m=-1/10$]{
\includegraphics[height=44mm]{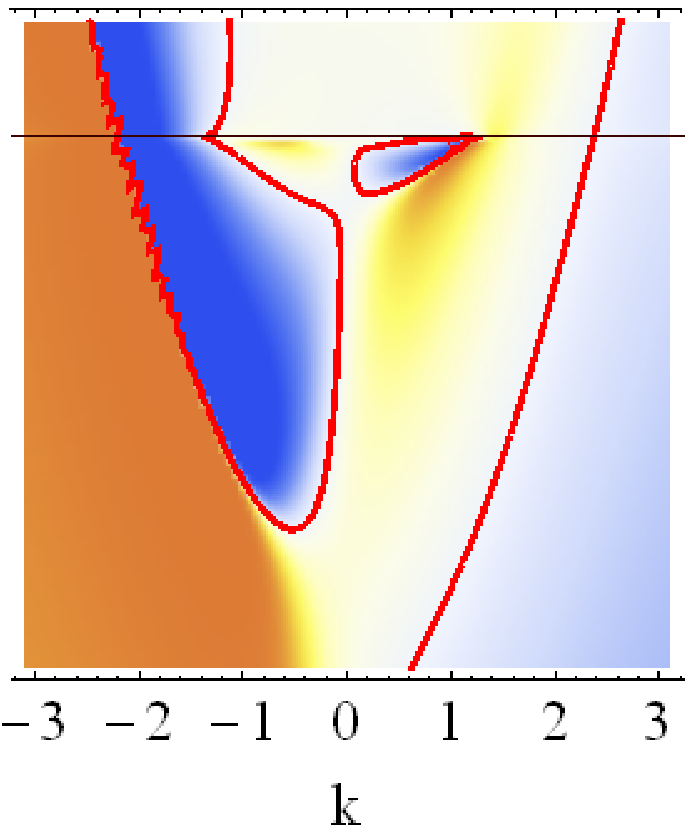}
\put(0,30)
{\includegraphics[height=30mm]{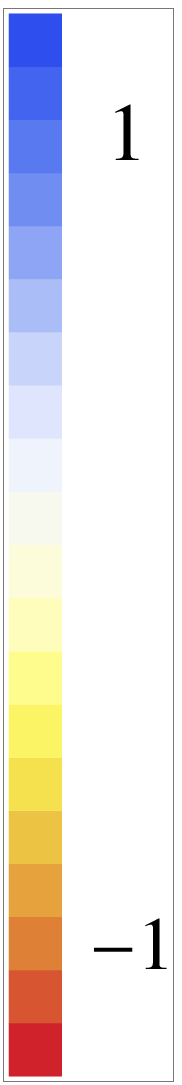}
}}
    \caption{Density plot  of $\Re G^{-1}$.
The zeros of the inverse Green's function represent the quasi-particle dispersion and are depicted by the red line (the wavy line indicates discontinuities).
Left panel: $m=0$. Right panel: $m=-0.1$. The dispersion crosses through the
Fermi wavevector at zero energy.  (The inset: lower energy range is shown. As in fig.\ref{pic_fs}-b, there is a dispersion branch with a different effective mass $m^*$)}
\label{pic_g11re}
\end{center}\end{figure}

Finally it is interesting to find a dispersion for the quasi-particle: the line where the real part of the inverse G-function is zero, $\Re\, G^{-1} = 0$. Since the  imaginary part is zero only at $\om=0$, the fermi surface is the intersection of the two lines. The results are illustrated in fig. \ref{pic_g11re}. The case of a massless fermion is particularly interesting, we see that the dispersion relation actually follows a parabolic form $\sqrt{p^2+ ( m^{*})^2}$ showing that the holographic fermion running around the doped black hole has developed a finite effective mass $m^{*}$. The dispersion at the Fermi surface is very similar to the linear dispersion of chiral fermion.

Summarizing the main points:
\begin{itemize}
\item [(a)]The Fermi Surface is rotationally invariant and has a single non-degenerate fermionic excitation at every momenta,

\item [(b)] The spin of the coherent excitations lies parallel to the momentum as shown in figure \ref{pic_fs}, giving rise to chiral quasi-particle interpretation,

\item [(c)]  The incoherent background is generated by a spin-flip of a coherent chiral quasiparticle.
\end{itemize}

\section{Discussion}

The possible application of holographic methods to condensed matter physics AdSCMT is based in part on a dream of a deep universality: the idea that the scale-invariance of quantum criticality in metals might enjoy the same level of universality seen in statistical physics.

Nevertheless, there is still a huge gulf to be crossed. From a String theory perspective, there is still a need to show that semi-classical gravity metric used in the theories emerges as a consistent truncation of string theory on a certain ``brane'' configuration. From a condensed matter perspective, we lack a
systematic method to constructing an AdS dual: one encouraging
direction may be to  map the renormalization group flows of the
quantum theory onto a higher dimension \cite{Lee:2012}.

Against this backdrop, the field has taken a more pragmatic approach of simply
exploring the holographic consequences of anti-de-Sitter space,
assuming that all is well. One can not fail to be impressed by the
discovery that a charged black hole nucleates a strange metal on its
surface, with properties that bear remarkable  similarities to
condensed matter systems: the emergence of
a critical Fermi surface with quantum oscillations, the presence of
E/T scaling, Pomeranchuk instabilities \cite{Pomeranchuk} and even the
phase diagram of type II superconductors \cite{type2}.

Our interest in the field was sparked by a naive impression that
progress in holography resembles the history of
condensed matter physics ``running in reverse''!  Rather
than starting with the simple Pauli paramagnetic metal and building up to an
understanding of Fermi liquids and ultimately quantum criticality,
holography appears to start with the
critical Fermi surface, working  backwards to the most basic
elements of condensed matter physics. This led us to ask, whether one
calculate the most elementary property of all, the Pauli
susceptibility in response to a Zeeman splitting.

Our work has provided a interpretation of the holographic
Green's function as a momentum and frequency dependent reflection
coefficient of waves emitted by surface particles, reflected off the
horizon of the interior black hole.  We have also shown that the
excitations of the strange metal are intrinsically
chiral, with spins locked parallel to the momentum by a strong
spin-orbit coupling with no inversion symmetry. The situation is reminiscent to the surface of 3D topological insulator.

This observation means in fact, that there is no Pauli susceptibility
of the two dimensional holographic metal: the Fermi sea is already
severely polarized by the relativistic coupling between momentum and
spin. Indeed, from a physical  perspective, the strange
metallic behavior seen in these systems would appear to be a
consequence of soft charge or current fluctuations rather than spin
fluctuations. In recent work \cite{Faulkner:2009am}, Faulkner et al
have discovered that when a charge gap is introduced by
condensing a boson in the AdS bulk, the holographic metal develops
sharp Fermi-liquid-like quasiparticles. This is consistent with this
interpretation.

Spin plays a major role in the quantum criticality of condensed
matter. In many systems, the application of a field, via the Zeeman
splitting is the method of choice for tuning through criticality \cite{YRS,Ybal1}.
Clearly, this part of the physics is inaccessible to the current
approach. The absence of a  Zeeman-splitting in holographic metals
was first observed by adding a monopole charge to the black hole
\cite{Albash:2009wz,Basu:2009qz}.

Various authors have explored the possibility of introducing spin as an
additional quantum number.
 The simplest example is the "magnetically
charged" black hole, with an ``up'' and a ``down'' charge to simulate
the Zeeman splitting, coupling to the fermions via a ``minimal
coupling'' (a spin-dependent vector
potential)\cite{Iqbal:2010eh}. By construction, this
procedure does produce an explicit ``Zeeman'' splitting of the Fermi
surface, however the infra-red character of the problem, described by the
interior geometry of the gravity dual, is unchanged and the strange
metal physics of the ``up'' and ``down'' Fermi surfaces are
essentially unaffected by the magnetic field.

An alternative approach might be
to introduce the Zeeman term to the holographic metals by invoking  a
non-minimal coupling to electromagnetic field, akin to the
anomalous magnetic coupling of a neutron or proton.
A number of
recent papers have considered the effect of such terms in the absence
of a magnetic field, where they play the role of anomalous dipole
coupling terms\cite{dipole1, dipole2}. At strong coupling these terms
have been found to inject a gap into the fermionic spectrum
interpreted as a Mott gap\cite{dipole2}. However, when a monopole
charge is added to the black hole to generate a magnetic coupling to
these same terms, we find they do not generate a splitting of
the chiral Fermi surface, nor do they change the interior geometry of the
gravity dual.
The construction of a holographic metal with non-trivial spin physics
may require considering 3-dimensional holographic metals
projected out of 4+1 dimensional
gravity dual \cite{Kraus}, where the additional dimensionality permits
four-component fermionic fields with both left and right-handed
chiralities.
\\

{\bf Note:} Shortly before posting our paper, a related work by
Hertzog and Ren\cite{Herzog} appeared with results that compliment
those derived here.  These authors concentrate on the behavior of
gravity duals with a large black-hole charge and a non-zero fermion
mass, a limit where the holographic metal contains multiple Fermi
surfaces.  They find that in this limit, in addition to a Rashba
component $\bs \cdot (\hat z\times \bk )$ the dispersion of the
holographic metal also develops a quadratic spin-independent
dispersion reminiscent of more weakly spin-orbit coupled fermions. The
Rashba term $\bs \cdot (\hat z\times\bk)$ obtained by
Herzog and Ren is equivalent to the
helicity term $\bs\cdot \bk $ described in our paper
after a rotation of spin axes. In the limit of small
black-hole charge considered here, with a single Fermi surface,
the helicity term in the Hamiltonian entirely
dominates the spectrum.
\\

\noindent \textbf{Acknowledgments.}

We gratefully acknowledge discussions with Steve Gubser, Matt
Strassler, Scott Thomas and Yue Zhao. This work was supported by
National Science Foundation grant DMR 0907179.

\appendix

\section{Boundary terms}\label{boundary.terms}
The general idea of holographic renormalization \cite{Skenderis:2002wp} is to add boundary terms to the classical gravity action. This terms simply make sure that all sensible physical quantities are finite. Good examples of such quantities are the total energy of the bulk (mass of a black hole inside) and the entropy. Those have nothing to do with duality and in some cases were introduces long before it, for instance by Hawking in 70's to actually make sense of his famous black hole temperature calculation. In AdSCFT it is useful to think about stability of a given AdS solution.

The Dirac action for fermions  in the bulk is of the form
\be
S = i\int d^{d+1}x  \sqrt{g}\    \overline{\psi} (\g^\mu D_\mu - m) \psi
\ee
in the bulk  $\psi_+$ and $\psi_-$ are related through each other momenta, but the conjugate momenta for $\overline{\psi}_+$ is zero:
\be
\Pi_{+} = -\sqrt{\frac {g}{g_{rr}} }\psi_- \quad \text{but } \quad
\overline{\,\Pi}_{+} = 0
\ee
Which is unphysical since we expect both momenta to represent a physical degree of freedom. The naive way to fix it which turns out to be the correct one is to change the bulk action by symmetrizing the kinetic term: split the derivative in half. One is acting to the left (represented by the arrow) and another is to the right.
 \be
S \rar \frac{i}2\int d^{d+1}x  \sqrt{g}  \  \overline{\psi} (\g^\mu \overrightarrow{D}_\mu -\g^\mu \overleftarrow{D}_\mu - 2m) \psi
\ee
which is different from the original action by a boundary term
\begin{equation}\label{boundary.fermion}
\d S_{\rm bound} = \int_{\p}  d^dx  \ \sqrt{\frac {g}{g_{rr}} }\ \ \overline{\psi}_+ \psi_- +h.c.
\end{equation}
And we are back to equation (\ref{ferm:action.bnd}). We refer to \cite{Henningson:1998cd} for more details.


\end{document}